\documentclass[sigconf]{acmart}
\usepackage{multirow}
\usepackage{threeparttable}
\usepackage{subfig}                 
\usepackage{overpic}
\usepackage{algorithm,algorithmic}

\newcommand{\liang}[1]{{\color{black}{#1}}}
\newcommand{\etal}{\textit{et al.}}

\usepackage{enumitem}
\setlist{topsep=0pt, leftmargin=*}
\AtBeginDocument{%
  \providecommand\BibTeX{{%
    \normalfont B\kern-0.5em{\scshape i\kern-0.25em b}\kern-0.8em\TeX}}}

\setcopyright{acmcopyright}
\copyrightyear{2018}
\acmYear{2018}
\acmDOI{10.1145/1122445.1122456}

\acmConference[Woodstock '18]{Woodstock '18: ACM Symposium on Neural
  Gaze Detection}{June 03--05, 2018}{Woodstock, NY}
\acmBooktitle{Woodstock '18: ACM Symposium on Neural Gaze Detection,
  June 03--05, 2018, Woodstock, NY}
\acmPrice{15.00}
\acmISBN{978-1-4503-XXXX-X/18/06}

\begin{document}

\title{Multi-Epoch Learning for Deep Click-Through Rate Prediction Models}

\author{Zhaocheng Liu}
\authornote{Both authors contributed equally to this work.}
\email{lio.h.zen@gmail.com}
\affiliation{%
  \institution{KuaiShou Technology}
  \city{Beijing}
  \country{China}
}

\author{Zhongxiang Fan}
\authornotemark[1]
\email{fanzhongxiang@kuaishou.com}
\affiliation{%
  \institution{KuaiShou Technology}
  \city{Beijing}
  \country{China}
}

\author{Jian Liang}
\email{liangjian03@kuaishou.com}
\affiliation{%
  \institution{KuaiShou Technology}
  \city{Beijing}
  \country{China}
}

\author{Dongying Kong}
\email{kongdongying@kuaishou.com}
\affiliation{%
  \institution{KuaiShou Technology}
  \city{Beijing}
  \country{China}
}

\author{Han Li}
\email{lihan08@kuaishou.com}
\affiliation{%
  \institution{KuaiShou Technology}
  \city{Beijing}
  \country{China}
}

\renewcommand{\shortauthors}{Trovato and Tobin, et al.}

\begin{abstract}
The one-epoch overfitting phenomenon has been widely observed in industrial Click-Through Rate (CTR) applications, where the model performance experiences a significant degradation at the beginning of the second epoch. 
Recent advances try to understand the underlying factors behind this phenomenon through extensive experiments.
However, it is still unknown whether a multi-epoch training paradigm could achieve better results, as the best performance is usually achieved by one-epoch training.
In this paper, we hypothesize that the emergence of this phenomenon may be attributed to the susceptibility of the embedding layer to overfitting, which can stem from the high-dimensional sparsity of data.
To maintain feature sparsity while simultaneously avoiding overfitting of embeddings, we propose a novel Multi-Epoch learning with Data Augmentation (MEDA), which can be directly applied to most deep CTR models.
MEDA achieves data augmentation by reinitializing the embedding layer in each epoch, thereby avoiding embedding overfitting and simultaneously improving convergence.
To our best knowledge, MEDA is the first multi-epoch training paradigm designed for deep CTR prediction models.
We conduct extensive experiments on several public datasets, and the effectiveness of our proposed MEDA is fully verified.
Notably, the results show that MEDA can significantly outperform the conventional one-epoch training.
Besides, MEDA has exhibited significant benefits in a real-world scene on Kuaishou.
\end{abstract}




\maketitle

\section{Introduction}
\label{section:intro}
Click-through rate (CTR) prediction is crucial for recommendations and online advertising.
In recent years, various deep learning-based CTR models \cite{cheng2016wide,qu2016product,guo2017deepfm,yu2020deep,zhou2018deep,zhou2019deep,pi2019practice,li2022adversarial} have emerged. 
In industrial CTR applications, the \textbf{one-epoch overfitting phenomenon} \cite{zhou2018deep,zhang2022towards} has been widely observed, which refers to a significant degradation in model performance at the beginning of the second epoch.
Unfortunately, most research \cite{belkin2019reconciling,salman2019overfitting,zhou2021over,zhang2021understanding,arpit2017closer} on the overfitting problem of Deep Neural Networks (DNNs) has been focused outside the CTR prediction field.
Moreover, to our best knowledge, in addition to the CTR domain, only large language models trained through supervised fine-tuning have reported similar one-epoch overfitting phenomena \cite{ouyang2022training}.
Recent advances \cite{zhang2022towards} in deep CTR prediction try to understand the underlying factors behind such phenomenon through extensive experiments.
However, it is still unknown whether a multi-epoch training paradigm could achieve better results, as the best performance is usually achieved by training with only one epoch.

The primary characteristic of CTR prediction lies in the extremely high-dimensional sparse nature of data.
Specifically, deep CTR prediction models are trained on large-scale datasets containing billions of categorical features, but with most of the features having low occurrence frequencies \cite{jiang2019xdl,zhao2019aibox,zhao2020distributed}.
To handle these categorical features, most deep CTR prediction models share a similar Embedding and Multi-Layer Perceptron (MLP) architecture.
Specifically, they typically adopt an embedding layer \cite{zhang2016deep} at the front, followed by various types of MLP structures, with the embedding layer responsible for mapping the high-dimensional categorical features to low-dimensional vectors.
According to a recent empirical study \cite{zhang2022towards}, lower feature sparsity reduces the occurrence of the one-epoch overfitting phenomenon.
Based on this, we hypothesize that the emergence of the one-epoch overfitting phenomenon may be attributed to the susceptibility of the embedding layer to overfitting, which can stem from the inherent high-dimensional sparsity of categorical features.
Moreover, in practice, lower feature sparsity leads to worse model performance.
Overall, to answer whether multi-epoch training can yield better performance and improve sample utilization efficiency, we need to find an approach that can maintain sufficient feature sparsity while simultaneously avoiding the overfitting of embeddings.



\begin{figure}
    \centering
    \includegraphics[width=0.9\linewidth]{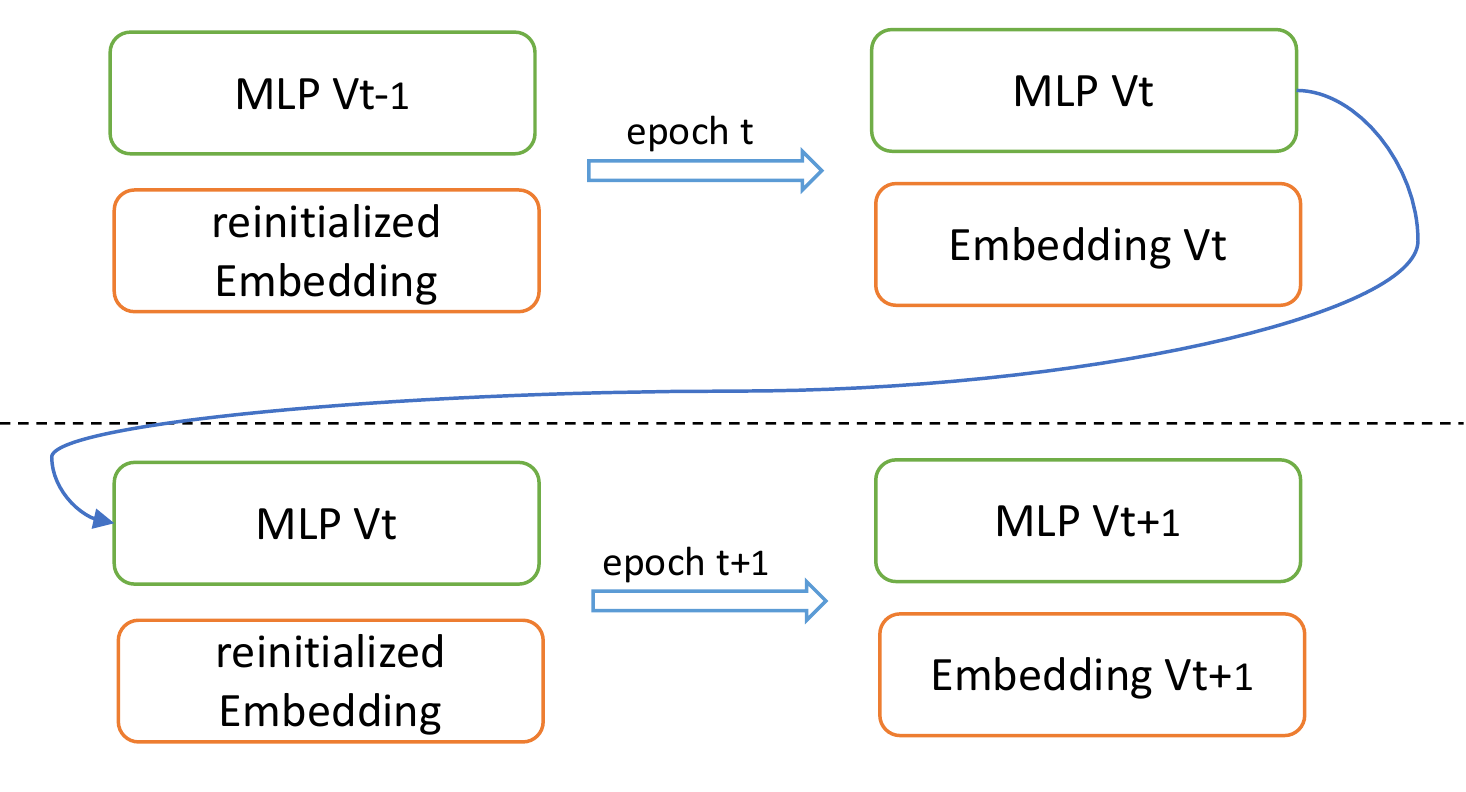}
    \caption{Our proposed Multi-Epoch learning with Data Augmentation (MEDA) reinitializes the embeddings in each epoch, which helps mitigate the issue of embedding overfitting while improving the convergence of the MLP.}
    \label{fig:multi_epoch_learning}
\end{figure}

In this paper, we adopt data augmentation to avoid overfitting of embeddings and  propose a novel Multi-Epoch learning with Data Augmentation (\textbf{MEDA}).
MEDA achieves data augmentation by reinitializing the embedding layer in each epoch, thereby avoiding overfitting of embeddings and simultaneously enhancing the convergence of the MLP.
Specifically, as shown in Figure \ref{fig:multi_epoch_learning}, the embedding component is reinitialized at the beginning of each training epoch and learns representations from the data anew.
Therefore, after undergoing multi-epoch training using MEDA, the embedding is in fact treated as a one-epoch training, thereby mitigating the occurrence of overfitting.
On the other hand, the MLP component follows regular multi-epoch training, where it iteratively updates its parameters with the augmented embeddings across multiple epochs.
This process enables the MLP to achieve enhanced convergence results.
This is because the MLP can not only learn more deeply over multiple epochs, but also come to understand that the real significance lies in the id itself, not the corresponding embedding.
To our best knowledge, MEDA is the first multi-epoch training paradigm designed for deep CTR prediction models.

To verify the effectiveness of our proposed MEDA, we conduct extensive experiments on several publicly available datasets.
Firstly, the second-epoch performance of training deep CTR models under MEDA outperforms the conventional one-epoch training by a significant margin.
Secondly, by comparing the metrics changes in multi-epoch training with and without MEDA, and exploring how many epochs are needed under less data to achieve the effect of one-epoch training with complete data, the effectiveness of data augmentation in MEDA has been fully validated. 
Thirdly, our ablation study reveals that as the number of epochs increases, the test AUC shows a steady improvement.
This allows for the flexibility to stop at any desired epoch based on the training cost, making it highly user-friendly in practical applications.
Lastly, MEDA has already been deployed in a real-world CTR prediction scenario on Kuaishou \footnote{\url{https://www.kuaishou.com/new-reco}} and demonstrated significant benefits in online A/B testing.

\section{RELATED WORK}
Experiments by \liang{Zhang \etal} \cite{zhang2022towards} show that reducing the sparsity to a sufficiently low level eliminated the one-epoch overfitting phenomenon.
This inspired us to focus our research on how to maintain data sparsity while avoiding embedding overfitting during multi-epoch training.
Applying our proposed MEDA to deep CTR models achieves significantly improved results.
MEDA is also relatively insensitive to the number of epochs, allowing flexible choice of training epochs based on cost.
In addition, \liang{Long \etal} \cite{ouyang2022training} noted that large language models also exhibit analogous one-epoch overfitting when conducting supervised fine-tuning.
However, they observed that a moderate degree of overfitting is, in fact, beneficial.
In our framework, this corresponds to appropriate embedding overfitting, which we leave to future work.

\section{THE PROPOSED APPROACH}
In this Section, we first introduce the Embedding\&MLP architecture followed by most popular deep CTR prediction models \cite{cheng2016wide,qu2016product,guo2017deepfm,yu2020deep,zhou2018deep,zhou2019deep,pi2019practice,li2022adversarial}.
Then we detail our proposed MEDA.

\subsection{The Embedding\&MLP Architecture}
\par\textbf{Input features.}
The features used in the deep CTR model include:
(1) Item profile: Item id and its side information (e.g., brand
id, shop id, category id, etc.).
(2) User profile: User id, age, gender, and income level.
(3) Long- and Short-Term User Behavior: For each user $u \in \mathcal{U}
$, there are several historical behavior sequences with different behavior types (e.g., impression, click, conversion, payment, etc.) and time windows representing long-term constant interests and short-term temporal needs.
In general, the raw features are first processed through feature discretization \cite{liu2020empirical} and feature selection \cite{guo2022lpfs} before being input into the model. Feature discretization converts numerical features into categorical features, resulting in most of the model's inputs being categorical features.
Unfortunately, in real-world industrial applications, categorical features are typically extremely high-dimensional sparse \cite{jiang2019xdl,zhao2019aibox,zhao2020distributed}, which is the primary characteristic of deep CTR prediction tasks and may significantly increase the risk of overfitting in deep CTR models \cite{zhang2022towards}.
\par\textbf{Embedding layer.}
To transform high-dimensional sparse features to low-dimensional dense representations, \liang{using an} embedding layer is a common operation for deep CTR models \cite{cheng2016wide,guo2017deepfm,yu2020deep,zhou2018deep,zhou2019deep,pi2019practice,li2022adversarial,zhang2022towards}.
For the i-th feature field $f_i$, let $\mathbf{E}^i = \left[e_1^i,...,e_j^i,...,e_{N^i}^i\right] \in \mathbb{R}^{D\times N_i}$ represent the i-th embedding dictionary, where $e_j^i$ is an embedding vector with dimensionality of $D$ and $N^i$ is the number of features in the i-th feature field.
The embedding representation of $f_i$ is a single embedding vector $e_j^i$ if $f_i$ is a one-hot vector with j-th element $f_i\left[j\right]=1$.
To be noted, $f_i$ also can be a multi-hot vector with $f_i\left[j\right]=1$ for $j \in \left\{i_1,i_2,...,i_k\right\}$.
In this case, as the embedding representation of $f_i$ is an embedding vector list $\left\{e_i^{i_1}, e_i^{i_2}, ..., e_i^{i_k}\right\}$, a pooling operation is adopted to get a fixed-length embedding vector.
\par\textbf{MLP.}
Given the concatenated dense representation vector, an MLP is employed to capture the nonlinear interaction among features \cite{liu2020dnn2lr}.
Numerous studies \cite{cheng2016wide,qu2016product,guo2017deepfm,yu2020deep} have concentrated on devising MLP architectures that excel in extracting pertinent information for tabular data.
It is worth noting that, given the rapid expansion of user historical behavior data in real-world industrial applications, careful attention must be paid to user behavior modeling during the design of MLPs.
The relevant methods \cite{zhou2018deep,zhou2019deep,pi2019practice,li2022adversarial} considering user behavior modeling focus on capturing the dynamic nature of user interests based on their historical behaviors \liang{which typically consist of lists of categorical features.}.
\begin{figure}
    \centering
    \includegraphics[width=\linewidth]{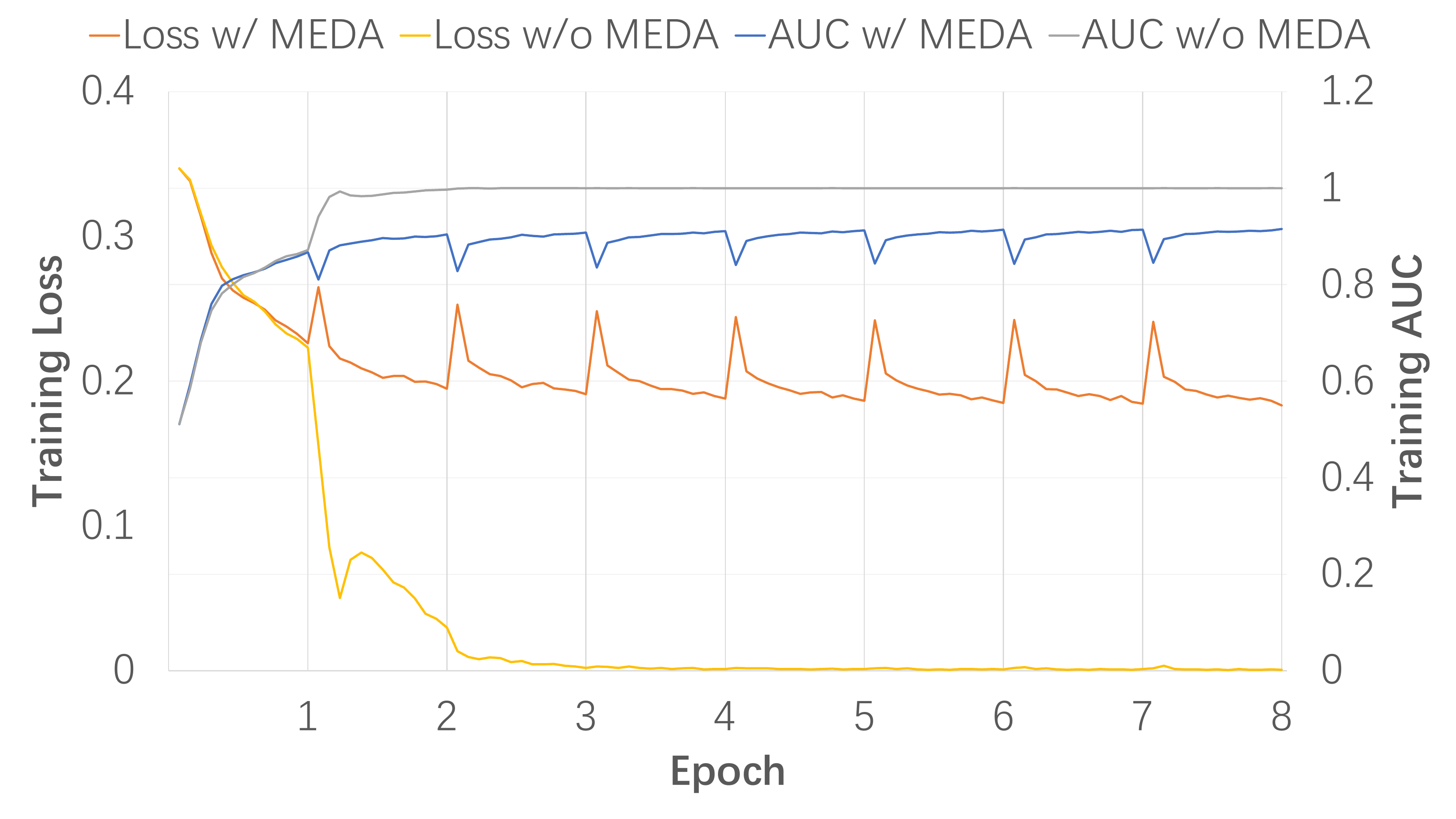}
    \caption{The training metric curves for DNN training 8 epochs on the Taobao dataset, comparing the results with and without the utilization of MEDA. }
    \label{fig:da_training}
\end{figure}
\subsection{The Proposed MEDA}
As introduced, for multi-epoch training, we propose MEDA to simultaneously maintain feature sparsity and avoid overfitting of embeddings. 
We show an overview of our proposed MEDA in Figure \ref{fig:multi_epoch_learning}; Algorithm \ref{alg_training} details the computational process of our MEDA framework. 
In addition, we consider several alternative structures, such as training normally in the first epoch, freezing embeddings from the second epoch, training the MLP several times, or freezing embeddings from the beginning and only training the MLP.
We find that other methods either do not converge or overfit.
In contrast, the proposed MEDA, a method with the meaning of data augmentation, achieves the expected results.
\begin{algorithm}
    \caption{Training process of MEDA}
    \label{alg_training}
    \begin{algorithmic}[1]
        \REQUIRE Training dataset $\mathcal{D}_{tr}$,  maximum training epoch $\mathop{n}$.
        \ENSURE MLP variable $\theta$ and embedding $\mathbf{E}$.
        \STATE random initialize the MLP variable $\theta_{0}$ and embedding $\mathbf{E}$.
        \FOR{$i=0$  to  $\mathop{n-1}$}
        \IF{$i>0$}
        \STATE random initialize $\mathbf{E}$,
        \ENDIF
        \STATE update $\theta_{i}$,$\mathbf{E}$ by one-epoch training,
        \STATE $\theta_{i+1}=\theta_{i}$,
        \ENDFOR
        \RETURN $\theta = \theta_{n}$, $\mathbf{E}$.
    \end{algorithmic}
\end{algorithm}
\section{EXPERIMENTS}
\begin{figure}
    \centering
    \includegraphics[width=\linewidth]{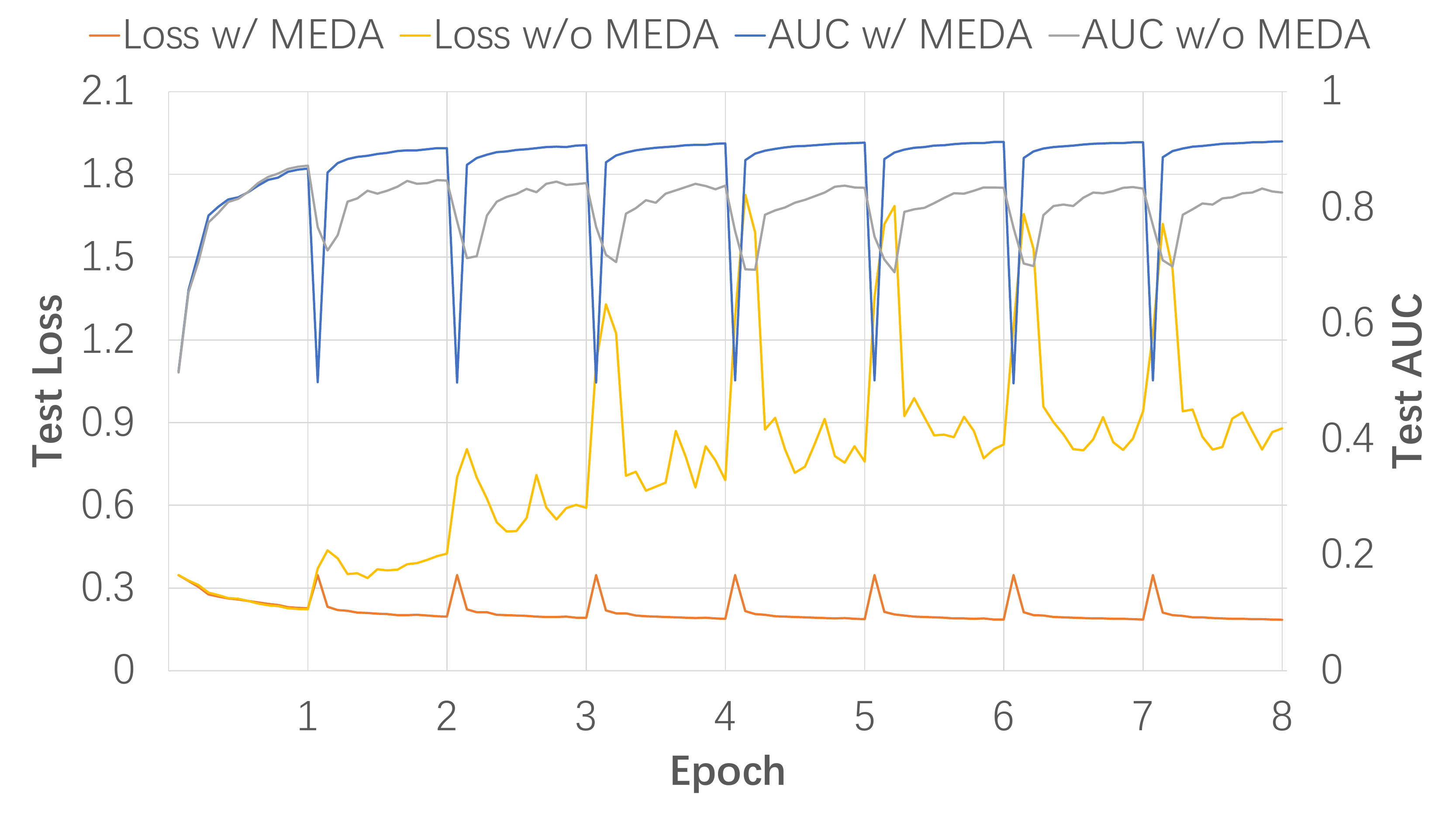}
    \caption{The test metric curves for DNN training 8 epochs on the Taobao dataset, comparing the results with and without the utilization of MEDA. }
    \label{fig:da_test}
\end{figure}
To verify the effectiveness of our proposed MEDA, we conduct extensive experiments on several publicly available datasets.
The experimental setup is detailed in Section 3.1, while the results and discussions are presented in Section 3.2.
Furthermore, Section 3.3 provides a thorough analysis of the impact of the number of epochs through ablation studies.
\subsection{Experimental setup}
\textbf{Datasets}. 
Two public datasets are used:
\par\textbf{Amazon dataset\footnote{\url{https://nijianmo.github.io/amazon/index.html}}}. It is a frequently used benchmark that consists of product reviews and metadata collected from Amazon \cite{ni2019justifying}. In our study, we specifically adopt the Books category of the Amazon dataset, which includes 51 million records, 1.5 million users, and 2.9 million items from 1252 categories. We regard the product reviews as click behaviors.
\par\textbf{Taobao dataset\footnote{\url{https://tianchi.aliyun.com/dataset/649}}}. It is a compilation of user behaviors from Taobao’s recommender system \cite{zhu2018learning}. This dataset encompasses 89 million records, 1 million users, and 4 million items from 9407 categories. In our analysis, we solely take into account the click behaviors of each user.
\\
\textbf{Baselines}. We apply our method on the following baselines:
\begin{itemize}
\item \textbf{DNN} is a base deep CTR model, consisting of an embedding layer and a feed-forward network with ReLU activation.
\item \textbf{DIN} \cite{zhou2018deep} proposes an attention mechanism to represent the user interests w.r.t. candidates.
\item \textbf{DIEN} \cite{zhou2019deep} uses GRU to model user interest evolution.
\item \textbf{MIMN} \cite{pi2019practice} proposes a memory network-based model to capture multiple channels of user interest drifting for long-term user behavior modeling.
\item \textbf{ADFM} \cite{li2022adversarial} proposes an adversarial filtering model on long-term user behavior sequences.
\end{itemize}
We denote above models as \textbf{Base}, and we incorporate them with the MEDA method as \textbf{Base+MEDA}.
\\
\textbf{Settings}.
All \textbf{Base} and \textbf{Base+MEDA} adhere to the optimal hyperparameters reported in their respective papers.
The optimization algorithm is Adam \cite{kingma2014adam} with a learning rate of 0.001.
We use Area under the curve (AUC) and binary cross-entropy loss as evaluation metrics.
\subsection{Performance evaluation}
Firstly, given that the number of training epochs is a hyperparameter, and multi-epoch training is costly in practice, the most economical and effective way to use MEDA is to train for 2 epochs.
We compare the performance of the second epoch (using MEDA) to evaluate whether it could achieve better results than the best performance obtained in single-epoch training for various commonly used deep CTR models. 
As shown in Table \ref{tab:second_epoch}, after applying MEDA, the second epoch performance of all methods significantly surpasses their best performance without MEDA (i.e., after a single epoch of training).

Secondly, we compared the changes in training and test loss, as well as AUC, of the same model when employing multi-epoch training with and without MEDA.
As illustrated in Figure \ref{fig:da_training} and Figure \ref{fig:da_test}, taking the DNN as an example, when MEDA is not employed, the one-epoch overfitting phenomenon occurs.
Simultaneously, when MEDA is employed, the training loss does not rapidly decrease from the second epoch onwards but rather decreases slowly, similar to how it performs when encountering new data.
Additionally, at the beginning of each epoch, the training loss slightly increases, resembling the performance when encountering new data from a different domain (resulting in numerous recognition errors).
Therefore, this experiment can provide some qualitative analytical support for MEDA to have the meaning of data augmentation.

Thirdly, to characterize the effectiveness of data augmentation in MEDA more quantitatively, we take the Taobao dataset as an example, randomly reducing the training data to 50\%, 25\% and 12.5\% of the original size, and conducting experiments to determine how many epochs of MEDA training (up to 16 epochs) are needed to recover the effect of one-epoch training with the complete data.
As shown in Table 2, MEDA can recover the effect of one-epoch training with complete data within 16 epochs, indicating that sample utilization efficiency has been significantly improved.
Moreover, the required number of epochs increases with the increasing complexity of the model.
Only ADFM is an exception, possibly because ADFM has a ultra-long behavior window, more interactions between id features, while the core of MEDA is to strengthen the importance of id information rather than the embedding itself by reinitializing the embedding. 

\begin{table}[]
\centering
\caption{The AUC performance on the Book and Taobao datasets, with Base indicating results obtained after one epoch of training, Base+MEDA representing results obtained after two epochs of training.}
\label{tab:second_epoch}
\scalebox{0.9}{
\begin{tabular}{c|c|c|c|c|c|c}
\hline
           & & DNN & DIN & DIEN & MIMN & ADFM \\
\hline
\multirow{3}{*}{Book} & Base & 0.8355 & 0.8477 & 0.8529 & 0.8686 & 0.8428 \\ \cline{2-7}
& Base+MEDA & \textbf{0.8450} & \textbf{0.8617} & \textbf{0.8602} & \textbf{0.8861} & \textbf{0.8507} \\ \cline{2-7}
& Improv. & +0.95\% & +1.4\% & +0.73\% & +1.75\% & +0.79\% \\ \hline
\multirow{3}{*}{Taobao} & Base & 0.8714 & 0.8804 & 0.9032 & 0.9392 & 0.9462 \\ \cline{2-7}
& Base+MEDA & \textbf{0.9034} & \textbf{0.9265} & \textbf{0.9262} & \textbf{0.9500} & \textbf{0.9568} \\ \cline{2-7}
& Improv. & +3.2\% & +4.61\% & +2.3\% & +1.08\% & +1.06\% \\ \hline

\end{tabular}
}
\end{table}

\begin{table}[]
\centering
\caption{The number of training epochs required for MEDA multi-epoch training on partial Taobao datasets to recover the effect of one-epoch training on the complete Taobao dataset and the corresponding Test AUC.}
\label{tab:how_many_epochs}
\scalebox{0.95}{
\begin{tabular}{c|c|c|c|c|c|c}
\hline
          &  & DNN & DIN & DIEN & MIMN & ADFM \\
\hline
\multirow{2}{*}{100\%} & \#Epochs & 1 & 1 & 1 & 1 & 1 \\ \cline{2-7}
 & Test AUC & 0.8714 & 0.8804 & 0.9032 & 0.9392 & 0.9462 \\ \hline
\multirow{2}{*}{50\%} & \#Epochs & \textbf{2} & \textbf{2} & \textbf{3} & \textbf{3} & \textbf{2} \\ \cline{2-7}
 & Test AUC & 0.8864 & 0.8989 & 0.9139 & 0.9444 & 0.9525 \\ \hline
\multirow{2}{*}{25\%} & \#Epochs & \textbf{4} & \textbf{3} & \textbf{6} & \textbf{13} & \textbf{2} \\ \cline{2-7}
 & Test AUC & 0.8802 & 0.8847 & 0.9048 & 0.9395 & 0.9466 \\ \hline
\multirow{2}{*}{12.5\%} & \#Epochs & \textbf{7} & \textbf{7} & \textbf{16} & 16 & \textbf{3} \\ \cline{2-7}
 & Test AUC & 0.8716 & 0.8844 & 0.9030 & 0.9287 & 0.9470 \\ \hline
\end{tabular}
}
\end{table}

\subsection{Ablation study}
\begin{figure}
    \centering
    \subfloat[Taobao]{\includegraphics[width=0.5\columnwidth]{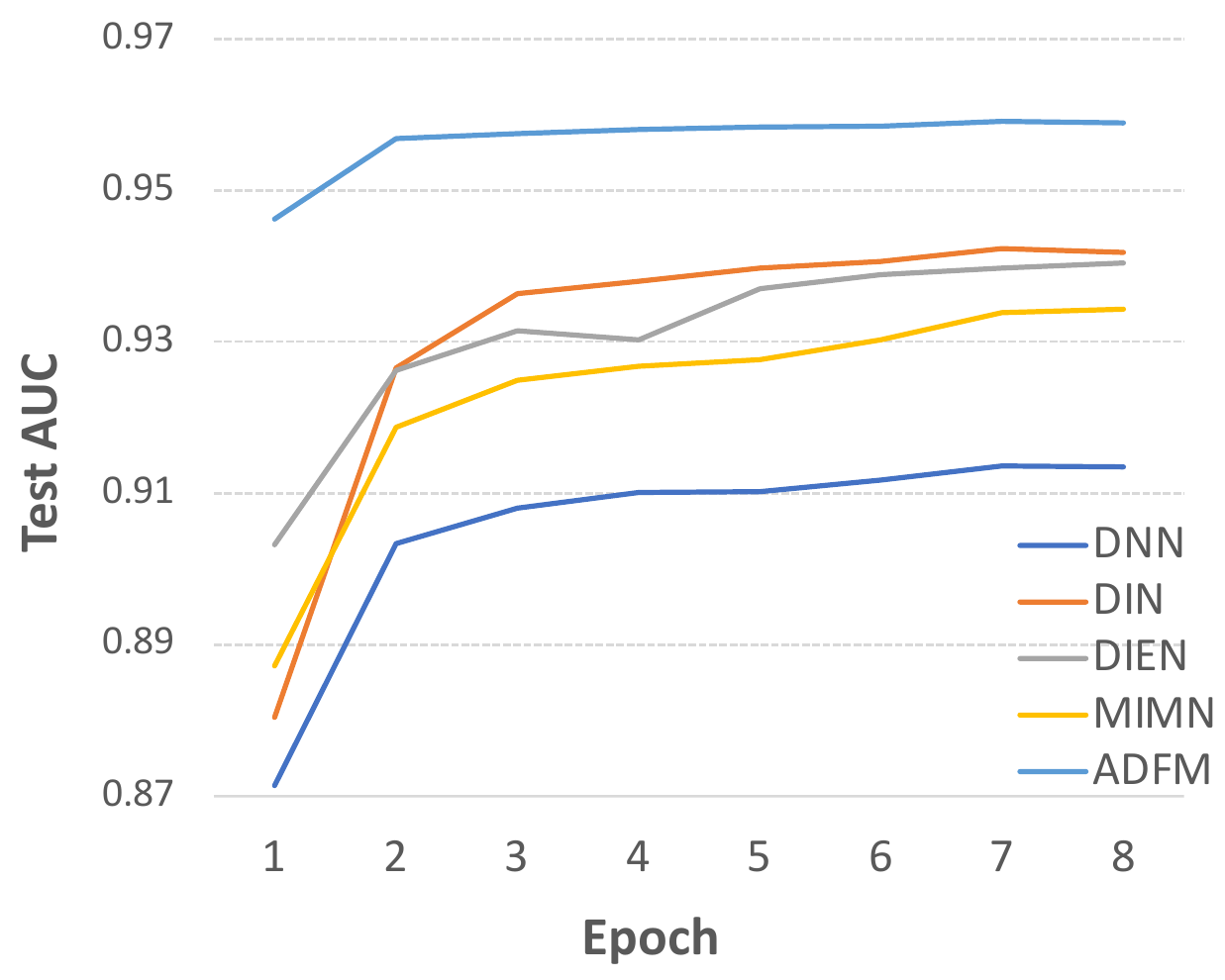}}
    \subfloat[Amazon]{\includegraphics[width=0.5\columnwidth]{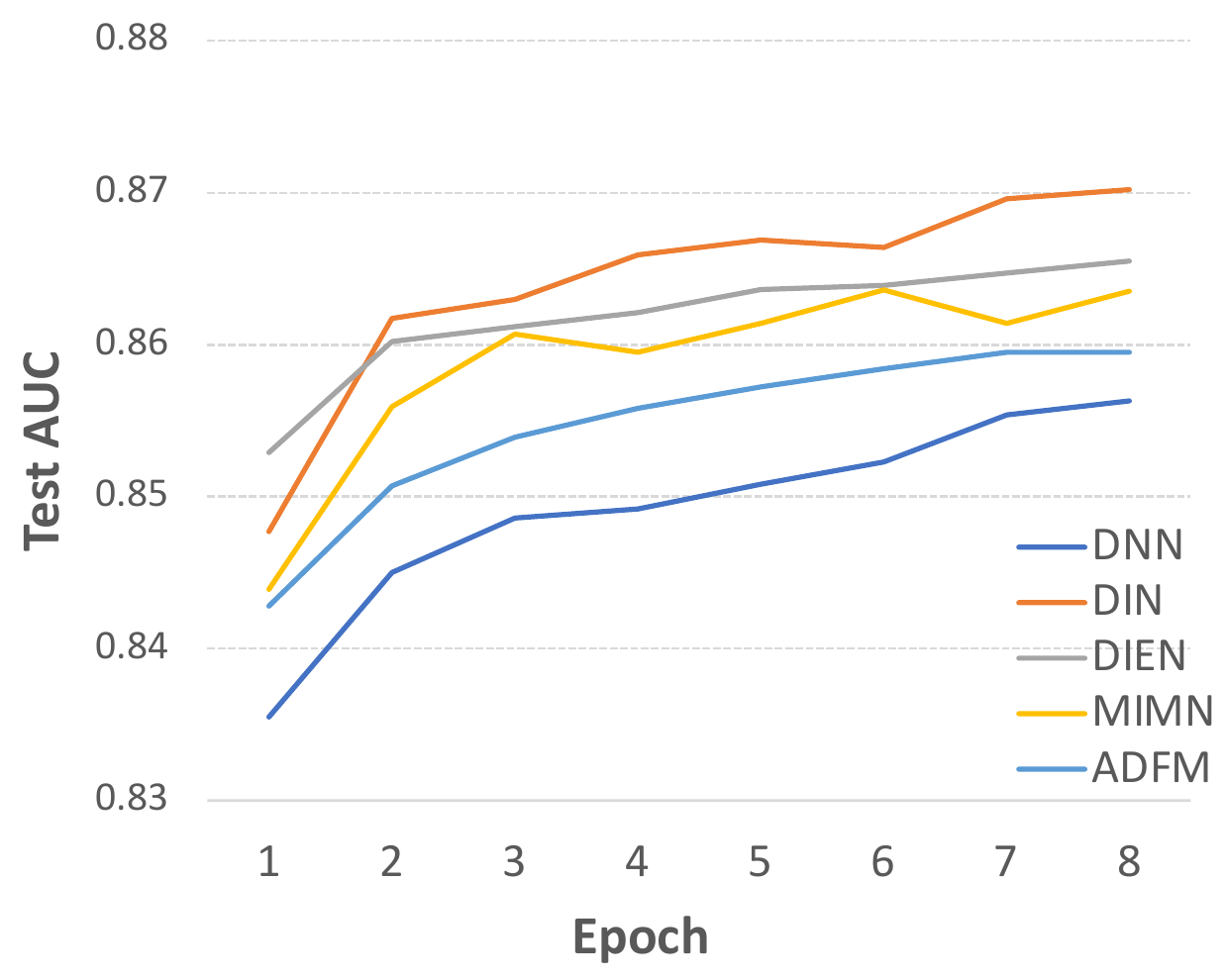}}
    \caption{The AUC curves of various models trained for 8 epochs on the Taobao dataset and the Amazon dataset.}
    \label{fig:ablation_study}
\end{figure}

To investigate the evolving trends of model performance across multiple epochs after incorporating MEDA, we conduct experiments on two public datasets.
Considering the typical size of industrial datasets, which can reach billions of samples, and the constraints imposed by computational resources, it is often only feasible to train models for 3-4 epochs.
Therefore, we trained all base+MEDA up to 8 epochs and recorded the test AUC at the end of each epoch.
As shown in Figure \ref{fig:ablation_study}, the majority of models exhibit a stable upward trend in performance as the number of epochs increased.
Therefore, it is feasible to stop at any epoch since the number of epochs does not significantly impact the outcome.
In practical applications, this allows users to determine the number of epochs based on training costs, making it highly user-friendly.

\subsection{Online result}
We conduct an online A/B experiment in a scene on Kuaishou, which lasted for 9 days. Compared with the previous state-of-the-art method used online, MEDA increased the AUC by 0.14\%, the cumulative revenue by 4.6\%, and the social welfare by 7.4\%.
To our best knowledge, this is the first approach that resolves the problem of overfitting in multi-epoch training of large-scale sparse models for advertising recommendations.
Moreover, achieving sufficient model performance for this scenario on Kuaishou requires training on approximately one month's worth of data. 
However, by employing the MEDA approach, comparable results can be achieved after training on just two weeks of data.
Therefore, utilizing MEDA can significantly reduce the required sample size to achieve equivalent results, leading to a significant reduction in training costs.

\section{Conclusion}
In this paper, we propose a novel  Multi-Epoch learning with
Data Augmentation.
MEDA achieves data augmentation by reinitializing the embedding layer in each epoch, thereby avoiding overfitting of embeddings and simultaneously enhancing the convergence of the MLP.
The experimental results show that MEDA effectively achieves the desired effect of data augmentation and multi-epoch learning can outperform
the conventional one-epoch training by a significant margin.
Moreover, MEDA has already been deployed in a real-world scenario
on Kuaishou, and has exhibited significant benefits during online
A/B Testing.

\balance
\bibliographystyle{ACM-Reference-Format}
\bibliography{sample-base}


\begin{thebibliography}{25}


\ifx \showCODEN    \undefined \def \showCODEN     #1{\unskip}     \fi
\ifx \showDOI      \undefined \def \showDOI       #1{#1}\fi
\ifx \showISBNx    \undefined \def \showISBNx     #1{\unskip}     \fi
\ifx \showISBNxiii \undefined \def \showISBNxiii  #1{\unskip}     \fi
\ifx \showISSN     \undefined \def \showISSN      #1{\unskip}     \fi
\ifx \showLCCN     \undefined \def \showLCCN      #1{\unskip}     \fi
\ifx \shownote     \undefined \def \shownote      #1{#1}          \fi
\ifx \showarticletitle \undefined \def \showarticletitle #1{#1}   \fi
\ifx \showURL      \undefined \def \showURL       {\relax}        \fi
\providecommand\bibfield[2]{#2}
\providecommand\bibinfo[2]{#2}
\providecommand\natexlab[1]{#1}
\providecommand\showeprint[2][]{arXiv:#2}

\bibitem[\protect\citeauthoryear{Arpit, Jastrz{\k{e}}bski, Ballas, Krueger,
  Bengio, Kanwal, Maharaj, Fischer, Courville, Bengio, et~al\mbox{.}}{Arpit
  et~al\mbox{.}}{2017}]%
        {arpit2017closer}
\bibfield{author}{\bibinfo{person}{Devansh Arpit},
  \bibinfo{person}{Stanis{\l}aw Jastrz{\k{e}}bski}, \bibinfo{person}{Nicolas
  Ballas}, \bibinfo{person}{David Krueger}, \bibinfo{person}{Emmanuel Bengio},
  \bibinfo{person}{Maxinder~S Kanwal}, \bibinfo{person}{Tegan Maharaj},
  \bibinfo{person}{Asja Fischer}, \bibinfo{person}{Aaron Courville},
  \bibinfo{person}{Yoshua Bengio}, {et~al\mbox{.}}}
  \bibinfo{year}{2017}\natexlab{}.
\newblock \showarticletitle{A closer look at memorization in deep networks}. In
  \bibinfo{booktitle}{\emph{International conference on machine learning}}.
  PMLR, \bibinfo{pages}{233--242}.
\newblock


\bibitem[\protect\citeauthoryear{Belkin, Hsu, Ma, and Mandal}{Belkin
  et~al\mbox{.}}{2019}]%
        {belkin2019reconciling}
\bibfield{author}{\bibinfo{person}{Mikhail Belkin}, \bibinfo{person}{Daniel
  Hsu}, \bibinfo{person}{Siyuan Ma}, {and} \bibinfo{person}{Soumik Mandal}.}
  \bibinfo{year}{2019}\natexlab{}.
\newblock \showarticletitle{Reconciling modern machine-learning practice and
  the classical bias--variance trade-off}.
\newblock \bibinfo{journal}{\emph{Proceedings of the National Academy of
  Sciences}} \bibinfo{volume}{116}, \bibinfo{number}{32}
  (\bibinfo{year}{2019}), \bibinfo{pages}{15849--15854}.
\newblock


\bibitem[\protect\citeauthoryear{Cheng, Koc, Harmsen, Shaked, Chandra, Aradhye,
  Anderson, Corrado, Chai, Ispir, et~al\mbox{.}}{Cheng et~al\mbox{.}}{2016}]%
        {cheng2016wide}
\bibfield{author}{\bibinfo{person}{Heng-Tze Cheng}, \bibinfo{person}{Levent
  Koc}, \bibinfo{person}{Jeremiah Harmsen}, \bibinfo{person}{Tal Shaked},
  \bibinfo{person}{Tushar Chandra}, \bibinfo{person}{Hrishi Aradhye},
  \bibinfo{person}{Glen Anderson}, \bibinfo{person}{Greg Corrado},
  \bibinfo{person}{Wei Chai}, \bibinfo{person}{Mustafa Ispir}, {et~al\mbox{.}}}
  \bibinfo{year}{2016}\natexlab{}.
\newblock \showarticletitle{Wide \& deep learning for recommender systems}. In
  \bibinfo{booktitle}{\emph{Proceedings of the 1st workshop on deep learning
  for recommender systems}}. \bibinfo{pages}{7--10}.
\newblock


\bibitem[\protect\citeauthoryear{Guo, Tang, Ye, Li, and He}{Guo
  et~al\mbox{.}}{2017}]%
        {guo2017deepfm}
\bibfield{author}{\bibinfo{person}{Huifeng Guo}, \bibinfo{person}{Ruiming
  Tang}, \bibinfo{person}{Yunming Ye}, \bibinfo{person}{Zhenguo Li}, {and}
  \bibinfo{person}{Xiuqiang He}.} \bibinfo{year}{2017}\natexlab{}.
\newblock \showarticletitle{DeepFM: a factorization-machine based neural
  network for CTR prediction}.
\newblock \bibinfo{journal}{\emph{arXiv preprint arXiv:1703.04247}}
  (\bibinfo{year}{2017}).
\newblock


\bibitem[\protect\citeauthoryear{Guo, Liu, Tan, Liao, Yang, Lei, Kong, Chen,
  and Liu}{Guo et~al\mbox{.}}{2022}]%
        {guo2022lpfs}
\bibfield{author}{\bibinfo{person}{Yi Guo}, \bibinfo{person}{Zhaocheng Liu},
  \bibinfo{person}{Jianchao Tan}, \bibinfo{person}{Chao Liao},
  \bibinfo{person}{Sen Yang}, \bibinfo{person}{Yuan Lei},
  \bibinfo{person}{Dongying Kong}, \bibinfo{person}{Zhi Chen}, {and}
  \bibinfo{person}{Ji Liu}.} \bibinfo{year}{2022}\natexlab{}.
\newblock \showarticletitle{LPFS: Learnable Polarizing Feature Selection for
  Click-Through Rate Prediction}.
\newblock \bibinfo{journal}{\emph{arXiv preprint arXiv:2206.00267}}
  (\bibinfo{year}{2022}).
\newblock


\bibitem[\protect\citeauthoryear{Jiang, Deng, Yi, Hu, Zhou, Zheng, Huang, Guo,
  Wang, Song, et~al\mbox{.}}{Jiang et~al\mbox{.}}{2019}]%
        {jiang2019xdl}
\bibfield{author}{\bibinfo{person}{Biye Jiang}, \bibinfo{person}{Chao Deng},
  \bibinfo{person}{Huimin Yi}, \bibinfo{person}{Zelin Hu},
  \bibinfo{person}{Guorui Zhou}, \bibinfo{person}{Yang Zheng},
  \bibinfo{person}{Sui Huang}, \bibinfo{person}{Xinyang Guo},
  \bibinfo{person}{Dongyue Wang}, \bibinfo{person}{Yue Song}, {et~al\mbox{.}}}
  \bibinfo{year}{2019}\natexlab{}.
\newblock \showarticletitle{Xdl: an industrial deep learning framework for
  high-dimensional sparse data}. In \bibinfo{booktitle}{\emph{Proceedings of
  the 1st International Workshop on Deep Learning Practice for High-Dimensional
  Sparse Data}}. \bibinfo{pages}{1--9}.
\newblock


\bibitem[\protect\citeauthoryear{Kingma and Ba}{Kingma and Ba}{2014}]%
        {kingma2014adam}
\bibfield{author}{\bibinfo{person}{Diederik~P Kingma} {and}
  \bibinfo{person}{Jimmy Ba}.} \bibinfo{year}{2014}\natexlab{}.
\newblock \showarticletitle{Adam: A method for stochastic optimization}.
\newblock \bibinfo{journal}{\emph{arXiv preprint arXiv:1412.6980}}
  (\bibinfo{year}{2014}).
\newblock


\bibitem[\protect\citeauthoryear{Li, Liang, Liu, and Zhang}{Li
  et~al\mbox{.}}{2022}]%
        {li2022adversarial}
\bibfield{author}{\bibinfo{person}{Xiaochen Li}, \bibinfo{person}{Jian Liang},
  \bibinfo{person}{Xialong Liu}, {and} \bibinfo{person}{Yu Zhang}.}
  \bibinfo{year}{2022}\natexlab{}.
\newblock \showarticletitle{Adversarial Filtering Modeling on Long-term User
  Behavior Sequences for Click-Through Rate Prediction}. In
  \bibinfo{booktitle}{\emph{Proceedings of the 45th International ACM SIGIR
  Conference on Research and Development in Information Retrieval}}.
  \bibinfo{pages}{1969--1973}.
\newblock


\bibitem[\protect\citeauthoryear{Liu, Liu, and Zhang}{Liu
  et~al\mbox{.}}{2020a}]%
        {liu2020empirical}
\bibfield{author}{\bibinfo{person}{Qiang Liu}, \bibinfo{person}{Zhaocheng Liu},
  {and} \bibinfo{person}{Haoli Zhang}.} \bibinfo{year}{2020}\natexlab{a}.
\newblock \showarticletitle{An empirical study on feature discretization}.
\newblock \bibinfo{journal}{\emph{arXiv preprint arXiv:2004.12602}}
  (\bibinfo{year}{2020}).
\newblock


\bibitem[\protect\citeauthoryear{Liu, Liu, Zhang, and Chen}{Liu
  et~al\mbox{.}}{2020b}]%
        {liu2020dnn2lr}
\bibfield{author}{\bibinfo{person}{Zhaocheng Liu}, \bibinfo{person}{Qiang Liu},
  \bibinfo{person}{Haoli Zhang}, {and} \bibinfo{person}{Yuntian Chen}.}
  \bibinfo{year}{2020}\natexlab{b}.
\newblock \showarticletitle{DNN2LR: Interpretation-inspired Feature Crossing
  for Real-world Tabular Data}.
\newblock \bibinfo{journal}{\emph{arXiv preprint arXiv:2008.09775}}
  (\bibinfo{year}{2020}).
\newblock


\bibitem[\protect\citeauthoryear{Ni, Li, and McAuley}{Ni et~al\mbox{.}}{2019}]%
        {ni2019justifying}
\bibfield{author}{\bibinfo{person}{Jianmo Ni}, \bibinfo{person}{Jiacheng Li},
  {and} \bibinfo{person}{Julian McAuley}.} \bibinfo{year}{2019}\natexlab{}.
\newblock \showarticletitle{Justifying recommendations using distantly-labeled
  reviews and fine-grained aspects}. In \bibinfo{booktitle}{\emph{Proceedings
  of the 2019 conference on empirical methods in natural language processing
  and the 9th international joint conference on natural language processing
  (EMNLP-IJCNLP)}}. \bibinfo{pages}{188--197}.
\newblock


\bibitem[\protect\citeauthoryear{Ouyang, Wu, Jiang, Almeida, Wainwright,
  Mishkin, Zhang, Agarwal, Slama, Ray, et~al\mbox{.}}{Ouyang
  et~al\mbox{.}}{2022}]%
        {ouyang2022training}
\bibfield{author}{\bibinfo{person}{Long Ouyang}, \bibinfo{person}{Jeffrey Wu},
  \bibinfo{person}{Xu Jiang}, \bibinfo{person}{Diogo Almeida},
  \bibinfo{person}{Carroll Wainwright}, \bibinfo{person}{Pamela Mishkin},
  \bibinfo{person}{Chong Zhang}, \bibinfo{person}{Sandhini Agarwal},
  \bibinfo{person}{Katarina Slama}, \bibinfo{person}{Alex Ray},
  {et~al\mbox{.}}} \bibinfo{year}{2022}\natexlab{}.
\newblock \showarticletitle{Training language models to follow instructions
  with human feedback}.
\newblock \bibinfo{journal}{\emph{Advances in Neural Information Processing
  Systems}}  \bibinfo{volume}{35} (\bibinfo{year}{2022}),
  \bibinfo{pages}{27730--27744}.
\newblock


\bibitem[\protect\citeauthoryear{Pi, Bian, Zhou, Zhu, and Gai}{Pi
  et~al\mbox{.}}{2019}]%
        {pi2019practice}
\bibfield{author}{\bibinfo{person}{Qi Pi}, \bibinfo{person}{Weijie Bian},
  \bibinfo{person}{Guorui Zhou}, \bibinfo{person}{Xiaoqiang Zhu}, {and}
  \bibinfo{person}{Kun Gai}.} \bibinfo{year}{2019}\natexlab{}.
\newblock \showarticletitle{Practice on long sequential user behavior modeling
  for click-through rate prediction}. In \bibinfo{booktitle}{\emph{Proceedings
  of the 25th ACM SIGKDD International Conference on Knowledge Discovery \&
  Data Mining}}. \bibinfo{pages}{2671--2679}.
\newblock


\bibitem[\protect\citeauthoryear{Qu, Cai, Ren, Zhang, Yu, Wen, and Wang}{Qu
  et~al\mbox{.}}{2016}]%
        {qu2016product}
\bibfield{author}{\bibinfo{person}{Yanru Qu}, \bibinfo{person}{Han Cai},
  \bibinfo{person}{Kan Ren}, \bibinfo{person}{Weinan Zhang},
  \bibinfo{person}{Yong Yu}, \bibinfo{person}{Ying Wen}, {and}
  \bibinfo{person}{Jun Wang}.} \bibinfo{year}{2016}\natexlab{}.
\newblock \showarticletitle{Product-based neural networks for user response
  prediction}. In \bibinfo{booktitle}{\emph{2016 IEEE 16th international
  conference on data mining (ICDM)}}. IEEE, \bibinfo{pages}{1149--1154}.
\newblock


\bibitem[\protect\citeauthoryear{Salman and Liu}{Salman and Liu}{2019}]%
        {salman2019overfitting}
\bibfield{author}{\bibinfo{person}{Shaeke Salman} {and} \bibinfo{person}{Xiuwen
  Liu}.} \bibinfo{year}{2019}\natexlab{}.
\newblock \showarticletitle{Overfitting mechanism and avoidance in deep neural
  networks}.
\newblock \bibinfo{journal}{\emph{arXiv preprint arXiv:1901.06566}}
  (\bibinfo{year}{2019}).
\newblock


\bibitem[\protect\citeauthoryear{Yu, Liu, Liu, Zhang, Wu, and Wang}{Yu
  et~al\mbox{.}}{2020}]%
        {yu2020deep}
\bibfield{author}{\bibinfo{person}{Feng Yu}, \bibinfo{person}{Zhaocheng Liu},
  \bibinfo{person}{Qiang Liu}, \bibinfo{person}{Haoli Zhang},
  \bibinfo{person}{Shu Wu}, {and} \bibinfo{person}{Liang Wang}.}
  \bibinfo{year}{2020}\natexlab{}.
\newblock \showarticletitle{Deep interaction machine: A simple but effective
  model for high-order feature interactions}. In
  \bibinfo{booktitle}{\emph{Proceedings of the 29th ACM International
  Conference on Information \& Knowledge Management}}.
  \bibinfo{pages}{2285--2288}.
\newblock


\bibitem[\protect\citeauthoryear{Zhang, Bengio, Hardt, Recht, and
  Vinyals}{Zhang et~al\mbox{.}}{2021}]%
        {zhang2021understanding}
\bibfield{author}{\bibinfo{person}{Chiyuan Zhang}, \bibinfo{person}{Samy
  Bengio}, \bibinfo{person}{Moritz Hardt}, \bibinfo{person}{Benjamin Recht},
  {and} \bibinfo{person}{Oriol Vinyals}.} \bibinfo{year}{2021}\natexlab{}.
\newblock \showarticletitle{Understanding deep learning (still) requires
  rethinking generalization}.
\newblock \bibinfo{journal}{\emph{Commun. ACM}} \bibinfo{volume}{64},
  \bibinfo{number}{3} (\bibinfo{year}{2021}), \bibinfo{pages}{107--115}.
\newblock


\bibitem[\protect\citeauthoryear{Zhang, Du, and Wang}{Zhang
  et~al\mbox{.}}{2016}]%
        {zhang2016deep}
\bibfield{author}{\bibinfo{person}{Weinan Zhang}, \bibinfo{person}{Tianming
  Du}, {and} \bibinfo{person}{Jun Wang}.} \bibinfo{year}{2016}\natexlab{}.
\newblock \showarticletitle{Deep Learning over Multi-field Categorical Data:
  --A Case Study on User Response Prediction}. In
  \bibinfo{booktitle}{\emph{Advances in Information Retrieval: 38th European
  Conference on IR Research, ECIR 2016, Padua, Italy, March 20--23, 2016.
  Proceedings 38}}. Springer, \bibinfo{pages}{45--57}.
\newblock


\bibitem[\protect\citeauthoryear{Zhang, Sheng, Zhang, Jiang, Han, Deng, and
  Zheng}{Zhang et~al\mbox{.}}{2022}]%
        {zhang2022towards}
\bibfield{author}{\bibinfo{person}{Zhao-Yu Zhang}, \bibinfo{person}{Xiang-Rong
  Sheng}, \bibinfo{person}{Yujing Zhang}, \bibinfo{person}{Biye Jiang},
  \bibinfo{person}{Shuguang Han}, \bibinfo{person}{Hongbo Deng}, {and}
  \bibinfo{person}{Bo Zheng}.} \bibinfo{year}{2022}\natexlab{}.
\newblock \showarticletitle{Towards Understanding the Overfitting Phenomenon of
  Deep Click-Through Rate Models}. In \bibinfo{booktitle}{\emph{Proceedings of
  the 31st ACM International Conference on Information \& Knowledge
  Management}}. \bibinfo{pages}{2671--2680}.
\newblock


\bibitem[\protect\citeauthoryear{Zhao, Xie, Jia, Qian, Ding, Sun, and Li}{Zhao
  et~al\mbox{.}}{2020}]%
        {zhao2020distributed}
\bibfield{author}{\bibinfo{person}{Weijie Zhao}, \bibinfo{person}{Deping Xie},
  \bibinfo{person}{Ronglai Jia}, \bibinfo{person}{Yulei Qian},
  \bibinfo{person}{Ruiquan Ding}, \bibinfo{person}{Mingming Sun}, {and}
  \bibinfo{person}{Ping Li}.} \bibinfo{year}{2020}\natexlab{}.
\newblock \showarticletitle{Distributed hierarchical gpu parameter server for
  massive scale deep learning ads systems}.
\newblock \bibinfo{journal}{\emph{Proceedings of Machine Learning and Systems}}
   \bibinfo{volume}{2} (\bibinfo{year}{2020}), \bibinfo{pages}{412--428}.
\newblock


\bibitem[\protect\citeauthoryear{Zhao, Zhang, Xie, Qian, Jia, and Li}{Zhao
  et~al\mbox{.}}{2019}]%
        {zhao2019aibox}
\bibfield{author}{\bibinfo{person}{Weijie Zhao}, \bibinfo{person}{Jingyuan
  Zhang}, \bibinfo{person}{Deping Xie}, \bibinfo{person}{Yulei Qian},
  \bibinfo{person}{Ronglai Jia}, {and} \bibinfo{person}{Ping Li}.}
  \bibinfo{year}{2019}\natexlab{}.
\newblock \showarticletitle{Aibox: Ctr prediction model training on a single
  node}. In \bibinfo{booktitle}{\emph{Proceedings of the 28th ACM International
  Conference on Information and Knowledge Management}}.
  \bibinfo{pages}{319--328}.
\newblock


\bibitem[\protect\citeauthoryear{Zhou, Mou, Fan, Pi, Bian, Zhou, Zhu, and
  Gai}{Zhou et~al\mbox{.}}{2019}]%
        {zhou2019deep}
\bibfield{author}{\bibinfo{person}{Guorui Zhou}, \bibinfo{person}{Na Mou},
  \bibinfo{person}{Ying Fan}, \bibinfo{person}{Qi Pi}, \bibinfo{person}{Weijie
  Bian}, \bibinfo{person}{Chang Zhou}, \bibinfo{person}{Xiaoqiang Zhu}, {and}
  \bibinfo{person}{Kun Gai}.} \bibinfo{year}{2019}\natexlab{}.
\newblock \showarticletitle{Deep interest evolution network for click-through
  rate prediction}. In \bibinfo{booktitle}{\emph{Proceedings of the AAAI
  conference on artificial intelligence}}, Vol.~\bibinfo{volume}{33}.
  \bibinfo{pages}{5941--5948}.
\newblock


\bibitem[\protect\citeauthoryear{Zhou, Zhu, Song, Fan, Zhu, Ma, Yan, Jin, Li,
  and Gai}{Zhou et~al\mbox{.}}{2018}]%
        {zhou2018deep}
\bibfield{author}{\bibinfo{person}{Guorui Zhou}, \bibinfo{person}{Xiaoqiang
  Zhu}, \bibinfo{person}{Chenru Song}, \bibinfo{person}{Ying Fan},
  \bibinfo{person}{Han Zhu}, \bibinfo{person}{Xiao Ma},
  \bibinfo{person}{Yanghui Yan}, \bibinfo{person}{Junqi Jin},
  \bibinfo{person}{Han Li}, {and} \bibinfo{person}{Kun Gai}.}
  \bibinfo{year}{2018}\natexlab{}.
\newblock \showarticletitle{Deep interest network for click-through rate
  prediction}. In \bibinfo{booktitle}{\emph{Proceedings of the 24th ACM SIGKDD
  international conference on knowledge discovery \& data mining}}.
  \bibinfo{pages}{1059--1068}.
\newblock


\bibitem[\protect\citeauthoryear{Zhou}{Zhou}{2021}]%
        {zhou2021over}
\bibfield{author}{\bibinfo{person}{Zhi-Hua Zhou}.}
  \bibinfo{year}{2021}\natexlab{}.
\newblock \showarticletitle{Why over-parameterization of deep neural networks
  does not overfit?}
\newblock \bibinfo{journal}{\emph{Science China Information Sciences}}
  \bibinfo{volume}{64} (\bibinfo{year}{2021}), \bibinfo{pages}{1--3}.
\newblock


\bibitem[\protect\citeauthoryear{Zhu, Li, Zhang, Li, He, Li, and Gai}{Zhu
  et~al\mbox{.}}{2018}]%
        {zhu2018learning}
\bibfield{author}{\bibinfo{person}{Han Zhu}, \bibinfo{person}{Xiang Li},
  \bibinfo{person}{Pengye Zhang}, \bibinfo{person}{Guozheng Li},
  \bibinfo{person}{Jie He}, \bibinfo{person}{Han Li}, {and}
  \bibinfo{person}{Kun Gai}.} \bibinfo{year}{2018}\natexlab{}.
\newblock \showarticletitle{Learning tree-based deep model for recommender
  systems}. In \bibinfo{booktitle}{\emph{Proceedings of the 24th ACM SIGKDD
  International Conference on Knowledge Discovery \& Data Mining}}.
  \bibinfo{pages}{1079--1088}.
\newblock


\end{thebibliography}

\end{document}